# Modeling of Sedimentation of Particles near Corrugated Surface by Boundary Singularity Method

by

Alex Povitsky

Professor, Department of Mechanical Engineering, The University of Akron, Akron OH USA, email: alex14@uakron.edu

**Abstract**

The velocity and trajectory of particle moving along the corrugated (rough) surface under action of gravity is obtained by meshless Boundary Singularity Method (BSM). This physical situation is found often in biological systems and microfluidic devices. The Stokes equations with no-slip boundary conditions are solved using the Green's function for Stokeslets. In the present study, the velocity of a moving particle is not known and becomes a part of the BSM solution. This requires an adjustment of the matrix of BSM linear system to include the unknown particle velocity and incorporate in the BSM the balance of hydrodynamic and gravity forces acting on the particle. Combination of regularization of Stokeslets and placement of Stokeslets outside flow domain was explored to ensure accuracy and stability of computations. Comparison has been made to prior published approximate analytical and experimental results to verify the effectiveness of this methodology to predict the trajectory of particle including its deviation from vertical trajectory and select the optimal set of computational parameters. The developed BSM methodology is applied to sedimentation of two spherical particles in proximity for which the analytical solution is not available.

**Introduction**

The flowfield with low value of Reynolds number, Re<<1, involving particle or micro-swimmer and corrugated (rough) surface is found often in biological systems and microfluidic devices.[1,2,3] In the current study, the particle is moving under action of gravity force acting in vertical direction. To ensure that the sum of hydrodynamic Stokes forces (shear stress and pressure) and gravity force exerted on the particle is equal to zero, the particle must have non-zero component of velocity perpendicular to the corrugated wall. That is, the particle could be either attracted or repelled by the rough wall along which it moves. In addition, the particle undergoes lateral displacement away from its vertical trajectory. The available analytical solution of Stokes equations[1,3] obtained under assumptions of small magnitude of roughness and small distance between a single particle and the wall compared to radius of spherical particle will be used for comparison to presented Boundary Singularity Method (BSM) results.

To model the flowfield, the meshless BSM is adopted. In general, the BSM offers a clear advantage in the broad range of geometries which can be handled. Using BSM, an arbitrary distance between particle and wall, arbitrary shapes of wall corrugation and shape of particle (or multiple particles) can be handled to include particles with irregular cross sections and multiple particles in proximity. As opposed to a regular BSM problem set-up, the velocity of moving particle is unknown in this study and is a part of BSM solution. This requires the adjustment of the matrix of BSM linear system to include the unknown particle velocity and incorporate zero balance of forces acting on the particle.

The boundary element methods (BEM) including the BEM variant named BSM, have validated and proven their efficiency in numerous fields of computational physics including micro- and nano- scale Stokes flows in channels, near particles, around fibers and their ensembles as shown in prior studies of the author[4,5,6,7,8,9]. The BSM is a special case of BEM based on linear combination of known fundamental solutions with singularities such as Stokeslets. The BSM has substantial advantages over traditional finite-volume (FV) and finite-difference (FD) mesh-based methods in computational fluid dynamics because BSM does not require cumbersome entire 3-D domain meshing. Instead, BSM requires placement of fundamental solutions (singularities) only at the boundaries of the considered geometry. This feature of BSM becomes especially critical for problems involving Stokes flows where the geometry of domain is complex including computing of flowfield about arbitrary-shaped particles in presence of rough surface.

The problem considered is challenging as when the sphere and the wall surface are close together, the singular behavior on one surface can adversely affect the integral taken over the other nearby surface[10] and, therefore the set-up of BSM have to be carefully selected. The BSM still suffers from issues related to the non-invertible or ill-conditioned matrix problem[11], therefore, it is important to maintain the moderate values of condition number. For this purpose, a combination of regularization of singularities and placement of singularities outside of flow domain will be explored in present study.

The analytical solution[1,3,12] of steady Stokes equations, to which the BSM results are compared in this study, is based on the first approximation of the flowfield by considering sedimentation of a sphere near a planar wall using a by-spherical coordinate representation. Such an approach can be used for a single spherical particle. As opposed to BSM, the approach is limited to small distance between the particle and wall and small magnitude of wall corrugations compared to the size of particle. The linear expansion about the first approximation is considered for small surface roughness amplitude with the effective slip velocity at the planar wall to mimic corrugated surface via a domain perturbation approach.

The goal of this paper is to adopt BSM to model the low Reynolds number flowfield of representative particle-to-corrugated wall combination. This will be done by solving the Stokes equations with no-slip boundary conditions, using the Green's function for the Stokeslet (single point force). Comparisons will be made to prior experimental and analytical results to verify the effectiveness of this methodology including prediction of deterministic lateral displacement of a single particle by flowfield near corrugation. Then, sedimentation of two particles near corrugated wall was modeled, to extend the BSM methodology to case in which analytical solution is not available.

The study is composed as follows. In Section 2 the geometric model, governing equations and numerical methodology is introduced. In Section 3, comparison of BSM solution with prior analytical solution for a particle at a fixed location with respect to corrugation wall is conducted and set of computational parameters is selected to ensure numerical accuracy. In Section 4, the BSM methodology is extended to a moving particle along the corrugated wall. The corrugations located with 45 degrees to the direction of gravity to compare the particle velocity and trajectory to available experimental and analytical results. In this section, the developed BSM methodology is applied to sedimentation of two spherical particles in proximity for which the analytical solution is not readily available.

## 2. The geometric model and numerical methodology

The considered geometric set-up includes a spherical particle in proximity to corrugated vertical wall (Fig. 1). For corrugated wall the roughness is presented as sinus function:

$$z = d + z_m \sin(2\pi x/w), \quad (2.1)$$

where x is the vertical direction along the wall, z is the normal direction to the wall into the flow domain, $d$ is the distance between the sphere and the smooth wall, $z_m$ is the magnitude of the wall roughness, and $w$ is the wavelength of periodical wall roughness. Variables are normalized by the radius of particle. The value of $x/w$ determines the local height of roughness.

A viscous steady flow with Re<1, which is caused by moving particle under action of gravity at terminal state, is represented by the system of simplified vector momentum equation (the Stokes equations) and a continuity equation, which can be written as follows:

$$\begin{cases} \nabla \cdot \vec{u} = 0 \\ -\nabla p + \mu \nabla^2 \vec{u} = 0. \end{cases} \quad (2.2)$$

The BSM involves two steps in its application: first, the substitution of the velocity, generated by Stokeslets with an unknown strength into the proper boundary conditions to construct a desired linear algebraic system of equations and the solution of this system; second, the use of the obtained strength of Stokeslets to compute the velocity and pressure distributions for entire domain. Three-dimensional velocity of the flow and the pressure induced by a point force, Stokeslet (the primary singularity of the Stokes flow), are given by[13]

$$u^{(k)} = \frac{1}{8\pi\mu}\left(\frac{F^{(k)}}{|\vec{r}|} + \frac{F^{(m)}r^{(m)}}{|\vec{r}|^3} r^{(k)}\right) \quad (2.3) \quad \text{and} \quad p = \frac{1}{4\pi}\left(\frac{F^{(m)}r^{(m)}}{|\vec{r}|^3}\right), \quad (2.4)$$

where $\vec{F}$ is the strength of an individual Stokeslet, $|\vec{r}|$ is the distance between the location of the Stokeslet and the collocation point, and superscripts, $k$ and $m$, denote components of the vectors, $\vec{F}$ and $\vec{r}$. The Einstein summation rule is used in above equations.

The collocation points, in which the no-slip and no-penetration boundary conditions should be satisfied, are located at the surface of sphere and at wall, see Fig. 1. The uniform distribution scheme of collocation points[4] is adopted in this study. This distribution assumes an approximately equal distance between collocation points over the spherical surface of particle (Fig. 1).

To avoid high condition number, $k$, of linear systems to be solved, singularities (Stokeslets) are located outside of computational domain as shown in Fig. 1. The Stokeslets are submerged underneath the wall surface and the sphere surface. This way the pole of the Green's function moves away from the boundary onto a surface exterior to the flow[13,14,15,16]. The submergence of singularities for Stokes equations and the optimum depth of submergence were discussed in[4,5] and references therein.

As another way to improve accuracy of BSM and reduce the condition number, the BSM uses the Stokeslets' regularization technique presented in[17,18]. BSM with submerged and/or regularized singularities are limited by their sensitivity either to the placement of singularities or to the value of regularization parameter. The value of regularization parameter $\varepsilon=\Delta s/4$, where $\Delta s$ is the distance between Stokeslets, is adopted as suggested[17]. Equations for velocity for Stokeslets with regularized singularities are as follows:

$$u^{(k)} = \frac{1}{8\pi\mu}\left(\frac{F^{(k)}(|\vec{r}|^2+2\varepsilon^2)}{|\vec{r}_\varepsilon|^3} + \frac{F^{(m)}r^{(m)}}{|\vec{r}_\varepsilon|^3}r^{(k)}\right), (2.5)$$

where $|\vec{r}_\varepsilon|^2 = |\vec{r}|^2 + \varepsilon^2$.

The above system (2.3) or its regularized version, (2.5), can be expressed in the matrix form for an arbitrary 3-D Stokes flow, as follows:

$$\vec{U} = M\,\vec{F}, (2.6)$$

where $\vec{U}$ is a velocity vector at collocation points (boundary conditions), $\vec{F}$ is a Stokes force vector with 3N unknowns, and M is a 3N x 3N matrix of above system. The Stokes force vector has the structure of $\vec{F} = \{F_1^x, F_1^y, F_1^z, ..., F_N^x, F_N^y, F_N^z\}$. For regular past implementation of BSM[4-9], the velocity vector at collocation points is equal to zero for no-slip and no-penetrating boundary conditions while the non-zero constant velocity is selected in far field. In the current study, the velocity vector, $\vec{U}$, is filled with unknown velocities, $u_x$, $u_y$ and $u_z$. As soon as the particle is non-deformable and non-rotating, its velocity components, $u_x$, $u_y$ and $u_z$, are the same for all collocation points. It is assumed that the particle has reached its terminal velocity. The presence of three new unknowns, $u_x$, $u_y$ and $u_z$, in the vector $\vec{U}$, the system (2.6) requires addition of three equations that can be presented as condition for the resultant force at terminal state of particle motion, which is equal to sum of Stokeslets:

$$\sum_{i=1}^{N} F_i^x = W\,;\ \sum_{i=1}^{N} F_i^y = 0\,;\ \sum_{i=1}^{N} F_i^z = 0, (2.7)$$

where W is the weight force directed in the $x$ direction and equal to one in normalized variables.

After addition of three equations (2.7) and three unknown velocities to system (2.6), its size becomes *(3N+3)x(3N+3)*. Numerical solution of above linear system (2.6) is performed using the MATLAB backslash linear algebraic operation, *F=U\M*[19].

Once the above linear system is solved, velocity at any point of the flowfield can be calculated by substituting Stokeslets, which are obtained by solving (2.6), either into (2.3) or into its regularized version, (2.5).

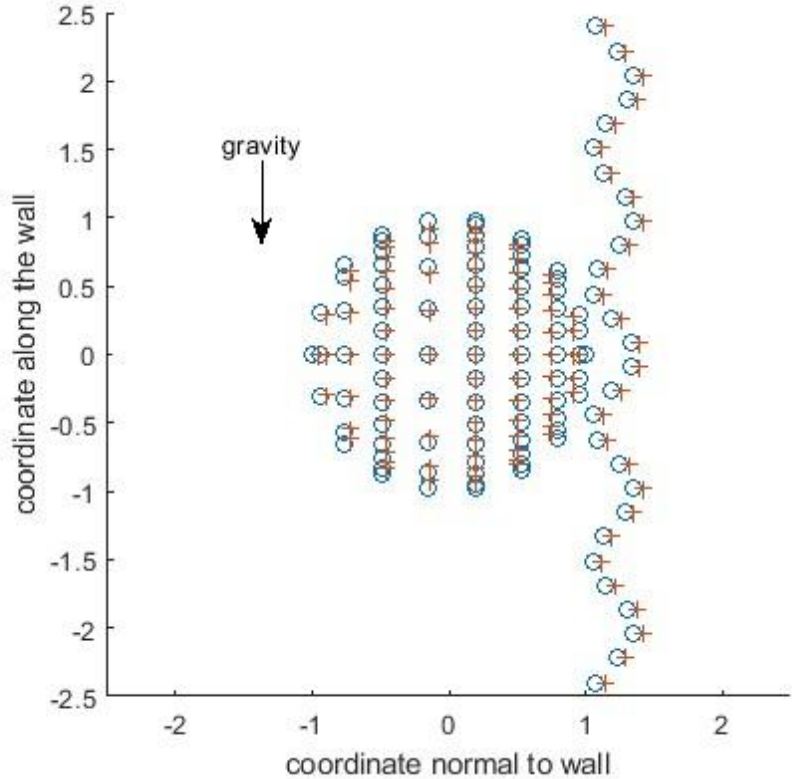

Figure 1: Spherical particle moving under action of gravity near corrugated vertical wall. Collocation points (o) and location of Stokeslets (+) are shown. Particle center is located at origin.

## 3. Computational parameters of BSM and comparison of numerical results with approximate analytical solution

In this section the particle is located at a fixed location with respect to corrugation (Fig. 1). The BSM solution is compared to prior analytical solution of flowfield around sedimented particle near corrugated wall[12]. The set of BSM computational parameters is selected in this section to ensure numerical accuracy and computational efficiency.

The geometric parameters listed below are normalized by the radius of particle. The given amplitude of corrugation, $z_m$, is equal to *0.15* and the corrugation wavelength, *w*, is equal to one. The distance between the sphere and the wall at the z direction, d, is equal to *0.2* so the minimum distance between the particle's surface and the wall is 0.05 and the distance between the wall and the center of particle is 1.2. The location of center of particle with respect to neutral level of corrugation is determined by parameter $\varphi$.

$$z = d + z_m \sin(\varphi + 2\pi x/w) \quad (3.1)$$

For $\varphi$ at its limit values, 0 or 1, the particle is facing the corrugated wall corresponding to the zero value of sinus, that is, the distance from particle to wall $z=d$. The values of $\varphi=0.25$ and $\varphi=0.75$ correspond to the maximum (see Fig. 1) and minimum distance between particle and wall, that is, the maximum and minimum of *sin* function in Eq. (3.1). At mid-values of phase parameter, $\varphi \sim 0.5$, the particle is facing the neutral level of corrugation, however, the gravity force is acting toward the corrugation in this case whereas for $\varphi$ at 0 or 1 the gravity force is acting toward the fluid.

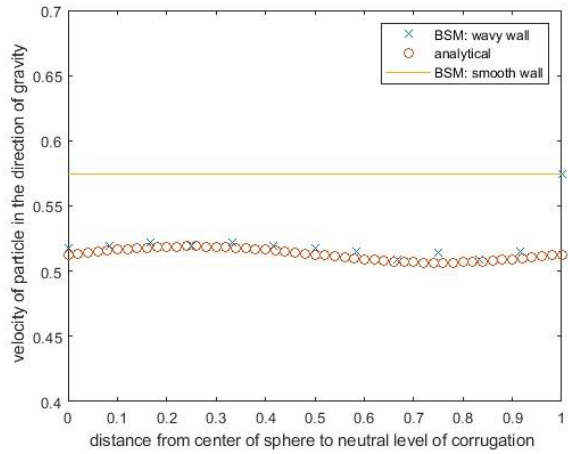

(a)

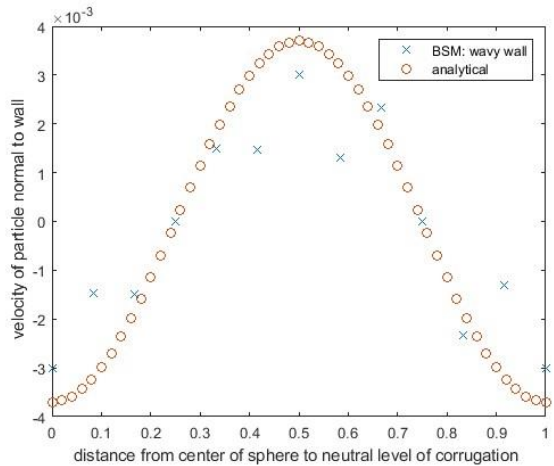

(b)

Figure 2: Velocity component (a) parallel to wall and (b) normal to wall as a function of parameter φ with comparison to analytical solution of Stokes equations[12] near corrugated wall. The number of Stokeslets and number of collocation points for the spherical particle is N=100.

The results are obtained by BSM (see Eqs (2.6-2.7)) using the 15 x 15 domain in *(x, y)* coordinates and shown in Figs. 2-3. The most accurate BSM results are obtained by regularized submerged Stokeslets for submergence depth of Stokeslets of 0.05 for spherical particle and 0.075 for the wall. The approximate distance between collocation points

$$\Delta = \sqrt{\frac{\pi a^2}{N}} \quad (3.2)$$

The Stokeslets located along the wall have halved distance between the neighboring Stokeslets in *x* and *z* directions, $\frac{\Delta}{2}$. The BSM solution for velocity component parallel to the wall does almost coincide with the analytical solution of Stokes equation near corrugated wall[12] (Fig. 2a). In Fig. 2a, BSM results are also presented for smooth vertical wall located at the same distance from the particle for comparison.

The velocity component normal to wall has a difference with the analytical solution of Stokes equations developed in Refs[1,3,12] for some of φ (Fig. 2b) especially those corresponding to particle facing the depth of the sin profile of roughness, that is, for mid-values of phase parameter. This situation has the positive value of normal component of velocity, that is, the particle is repelled off the wall using the coordinate system in Fig. 1. Note (compare Fig. 2a to Figs. 2b and 3) that the magnitude of normal component of velocity is $O(10^{-3})$ while the magnitude of component of velocity parallel to wall is $O(1)$; therefore, it is more difficult to maintain accuracy of velocity computations normal to wall. Further discussion aims at monitoring this difference and investigating the role of set-up parameters.

Results obtained by using 10x10 *x-y* domain are very close to those obtained by 15 x15 domain, therefore, the next results are presented for 10x10 domain. In Fig.3 results are presented for the following numbers of Stokeslets per sphere (a) N=150 and (b) N=200. Results for N=150 are somewhat improved compared to those for N=100 (compare Fig 2b to Fig. 3b) especially for the mid-range values of parameter φ. Results for N=200 show the deviation from analytic values. This can be explained by the dramatic increase of condition number, k, for N=200, in which case the values of k~$O(10^7)$. These values of condition number are much larger than those for N=100 (k,~650) and for N=150 (k~2300) that causes some loss of accuracy for N=200.

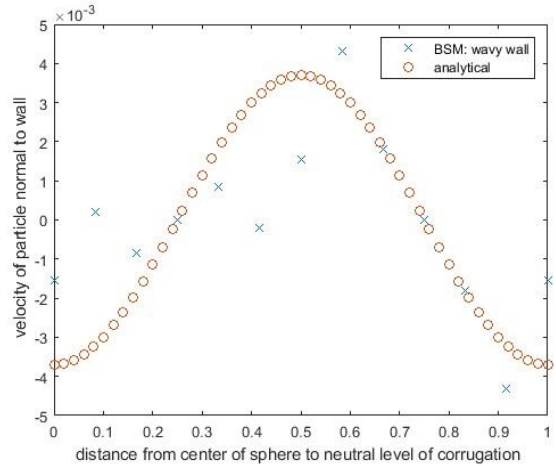

(a)

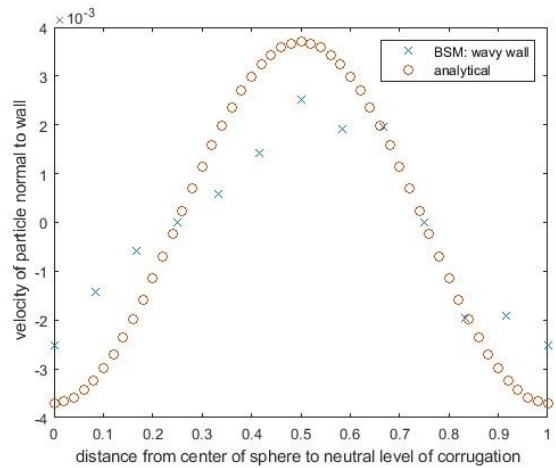

(b)

Figure 3: Velocity normal to wall for (a) N=200 and (b) N=150 as a function of parameter φ Velocity obtained by BSM is compared to analytical solution.

## 4. BSM computations for corrugations tilted with respect to gravity and for two particles

In this section, results of BSM computations are presented to determine trajectory of particle near the corrugated wall with tilted corrugations. The computational parameters are the same as those obtained in previous sections unless otherwise stated.

The authors compared theoretical and experimental results published in Ref[3] (Fig.3) against the BSM computations. The derivation of analytical solution of Stokes equations for prediction of roughness-induced velocity is presented in Ref[3], p. 7. The experimental methodology of imaging of the trajectory of the particle involved in sedimentation near 3-D-printed corrugated surface is presented in Ref.[3], p. 6. In these experiments using tweezers a particle is placed nearby the corrugated surface and then released. To minimize the effect of initial release/unsteadiness the particle is allowed to sediment ~10 times its radius before recording its trajectory[3]. The two 3-D printed corrugations of sinusoidal shape, Eq.(2.1), are tilted with the angle of 45 degrees with the vertical gravity force. The corrugation's wavelength is 6 mm. The particle's radius is 1.6 mm and the initial distance between the particle surface and the smooth wall is 0.2 mm.

The length of domain along the vertical coordinate is 3.5 wavelengths, $x/\lambda=3.5$, centered at $x=0$. A particle starts at *0.75 λ* upstream of the first corrugation and trajectory computing ends at *0.75 λ* downstream of the second corrugation. At first, the particle distance to the wall is equal to 0.2 mm as specified above. The size of domain in the *y* and *x* directions is 10*a* x 10*a*, where a is the particle radius. The domain is centered at the current particle location and moves together with the moving particle after each time step. Variation of the domain size from *8a x 8a to* 15*a* x 15*a* does not affect the obtained particle trajectory. Total of two hundred equal time steps, Δt, is taken to follow the particle's motion along the vertical *x* coordinate.

At each time step the flowfield around the particle is assumed to become relaxed to steady-state Stokes equations (2.2). The new values of velocity $u_x$, $u_y$ and $u_z$ are found by the solution of linear system (2.6). After computing of new velocity, new coordinates of particle are found by the following relations:

$$x_{new} = x_{old} + u_x \Delta t \;;\; y_{new} = y_{old} + u_y \Delta t \text{ and } z_{new} = z_{old} + u_z \Delta t \quad (4.1)$$

For each time step in a trajectory a large linear system (2.6) representing the BSM must be solved[20]. Results presented in Fig.4 compare BSM computations against published experimental and analytical results (see Ref[3], Fig. 3B). In Fig. 4, the distance in the normal direction from smooth wall is normalized by the corrugation wavelength, $z/\lambda$, and presented as a function of vertical coordinate along the wall, $x/\lambda$. By Fig. 4, experimental data points are located closer to BSM computations compared to analytical results in areas near minimum of $z/\lambda$. For areas near maximum of $z/\lambda$, the analytical and BSM computations are closer to each other compared to areas of minimum.

Results in Fig. 5 compare lateral displacements obtained by BSM computations against published experimental and analytical results (see Ref[3], Fig. 3C for sinusoidal corrugation). In Fig. 5, the lateral horizontal displacement is normalized by the corrugation wavelength, $y/\lambda$, and presented as a function of coordinate along the wall, $x/\lambda$. Results in Fig. 5 show that obtained BSM maximums of magnitude of lateral displacement, in terms of $y/\lambda$, are not as prominent as maximums obtained analytically and experimentally[3] but they still coincide in terms of their locations, $x/\lambda$, and distance between maximums.

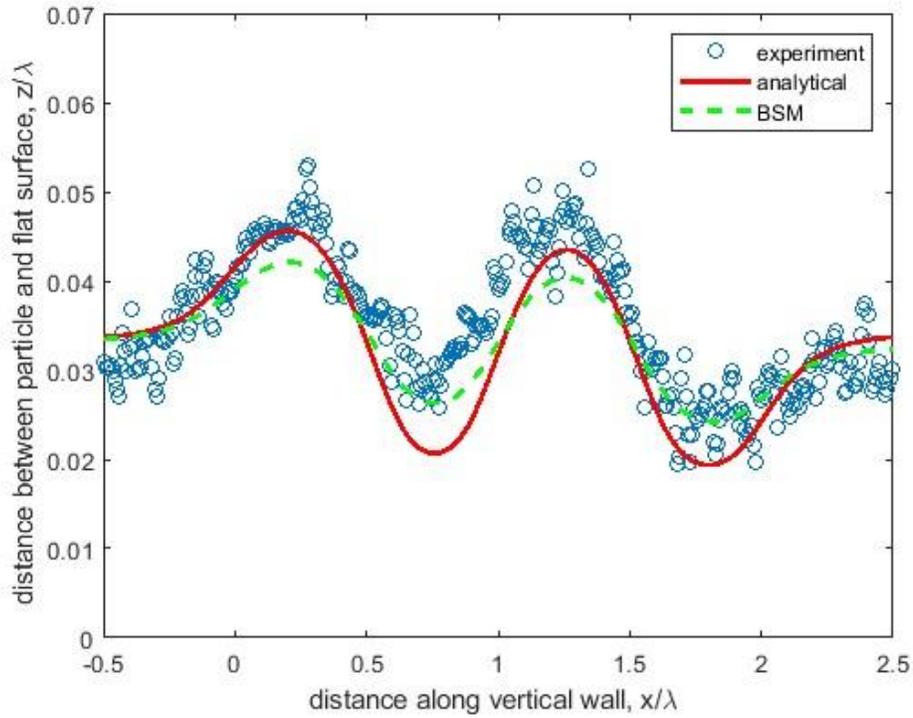

Figure 4: Comparison of the distance to wall obtained by BSM to published analytical and experimental results[3] for particle moving by action of gravity along tilted corrugations.

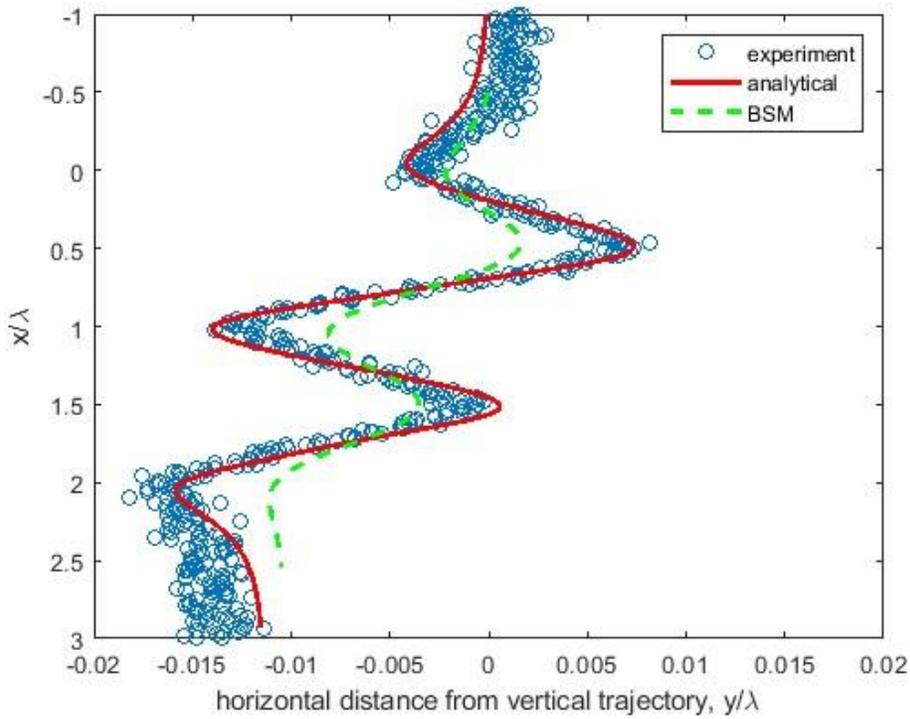

Figure 5: Comparison of the lateral displacement of particle obtained by BSM to published analytical and experimental results[3].

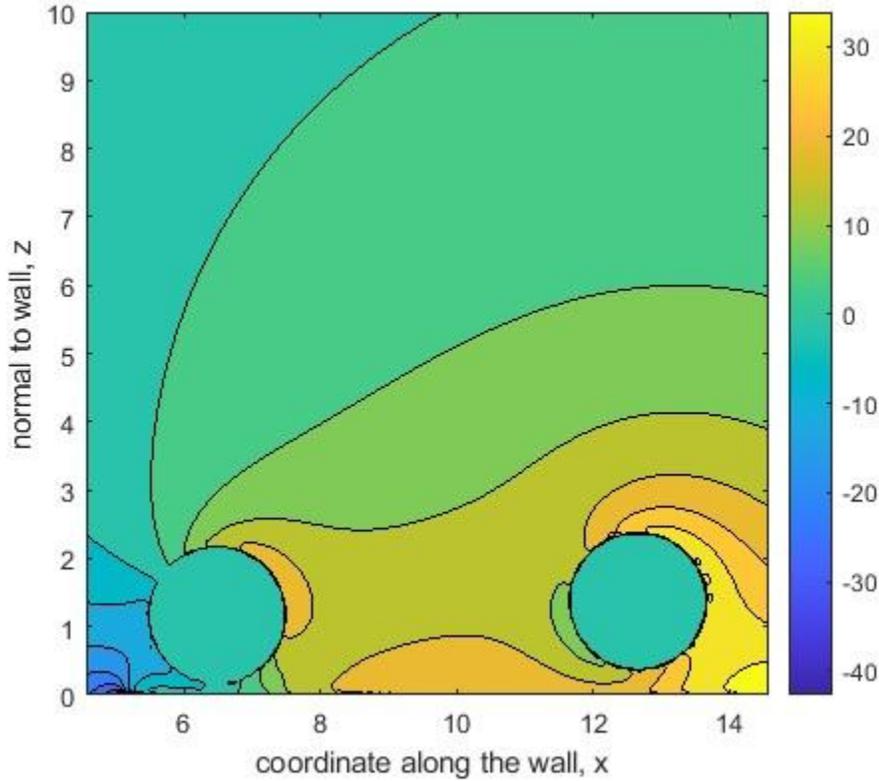

Fig. 6 Set-up of two particles near corrugated wall and distribution of static pressure.

Initially, the particles were located in proximity to the vertical wall with sinusoidal roughness (see above). The initial distance between centers of particles in vertical direction is equal to $5a$, where $a$ is the radius of particle. The set-up of two particles near corrugated wall and distribution of static pressure computed by Eq. (2.4) with modification for regularized Stokeslets[17] is shown in Fig. 6.

Trajectories are computed by BSM using the same distribution of Stokeslets per spherical particle as in above sections, N=150, with 200 time steps. In Fig. 7 trajectories of two particles moving along rough wall under action of gravity are depicted.

In Fig. 7a the deviation from vertical direction in the normal direction to wall is depicted. Recall that initial value of $z/\lambda$=0.2 mm/ 6 mm=0.033 for both particles. The first particle is pushed away from the wall by the moving second particle in its wake whereas the second particle is moving in closer proximity to wall compared to its initial distance from the wall. The deviation from vertical trajectories in horizontal direction is depicted in Fig. 7b. The first particle follows a path which is qualitatively similar to that of a single particle in Fig. 5, however, its deviation from vertical trajectory is less prominent compared to that for a single particle (Fig. 5) as its distance from surface roughness is larger. On the contrary, the deviation of the second particle is more than doubled compared to that of a single particle (Fig. 5) as the second particle passes closer to wall roughness.

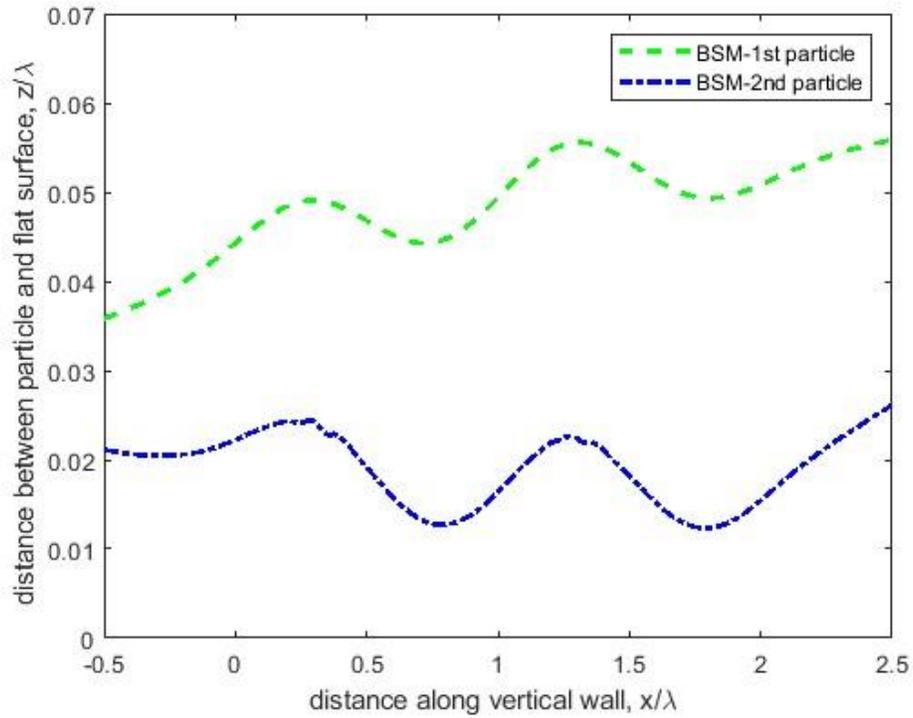

(a)

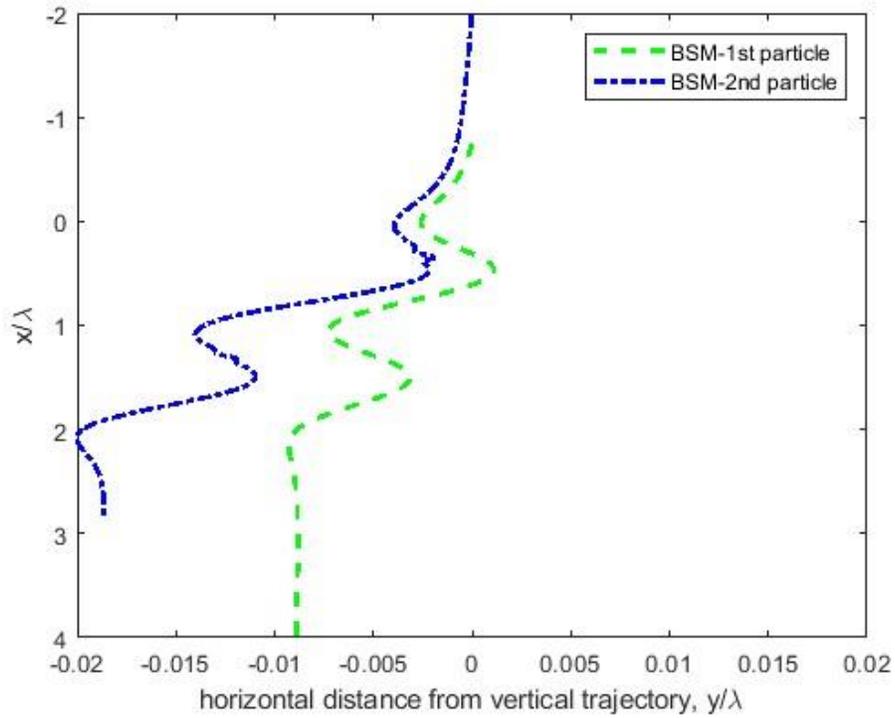

(b)

Figure 7: Trajectories of two particles moving along rough wall under action of gravity computed by BSM: (a) deviation from vertical direction in normal direction to wall and (b) deviation in horizontal direction.

**Conclusions**

The Boundary Singularity Method (BSM) is applied to a particle moving along the corrugated wall under the action of gravity to obtain its local velocity and trajectory. Modification of BSM linear system is implemented and verified to account for gravity force as a driver for the particle motion. Combination of regularization and submergence of Stokeslets is implemented to maintain the moderate value of condition number of BSM matrix to ensure accuracy of the solved linear system for the challenging case of closed proximity between the particle and wall. Obtained trajectory of particle including its lateral displacement does correspond to published experimental data and approximate analytical solution of Stokes equations.

Using the obtained set of parameters, the BSM methodology was extended to the case of two moving particles in proximity for which analytical solution is not applicable. The results show that particles deviate from trajectory of a single particle where the forward particle is moving farther away from vertical wall and the back particle is shifted closer to the wall. As a result, the horizontal deviation of the forward particle is smaller whereas the horizontal deviation of back particle is larger compared to that for a single particle. In the future research, the BSM will be extended to more complex situations of sedimentation in which analytical solution is not feasible, such as stochastic non-periodical wall roughness and clusters of particles shaped as either multiple spheres or Cassini ovals.

**Acknowledgements**

The author would like to thank the Visitors Program of Max Planck Institute for the Physics of Complex Systems (MPI) in Dresden, Germany, where this study started, to enable his research visit in Fall 2022 during his sabbatical leave from the University of Akron. The author acknowledges Dr. Christina Kurzthaler, Research Group Leader, MPI, for discussion and for pointing the author to her published analytical results for Stokes equations for comparison to presented BSM computations.
Matlab version R2022b is used in the current study.